\documentclass[12pt]{iopart}

\usepackage{epsfig,epsf}
\usepackage{graphicx}
\usepackage{epstopdf}
\newcommand{\beq}{\begin{equation}}
\newcommand{\eeq}{\end{equation}}
\def\eq#1{{(\ref{#1})}}

\newcommand{\as}{\alpha_S}

\newcommand{\lb}{\left(}
\newcommand{\rb}{\right)}

\begin{document}

%\title[Author guidelines for IOP journals in  \LaTeXe]{How to prepare and submit an article for 
%publication in an IOP journal using \LaTeXe}
\title[Heavy Ion Collisions at the LHC:
Last Call for Predictions, CERN, 2007]
{Hadron multiplicities at the LHC}
\author{D Kharzeev$^a$, E Levin$^b$ and M Nardi$^c$}

\address{
a) Department of Physics, Brookhaven National Laboratory,\\
Upton, New York 11973-5000, USA\\
b) HEP Department, School of Physics,\\
Raymond and Beverly Sackler Faculty of Exact Science,\\
Tel Aviv University, Tel Aviv 69978, Israel\\
c) INFN, Sezione di Torino,\\
via Giuria N.1, 10125 Torino, Italy
}
\ead{kharzeev@bnl.gov\\ leving@post.tau.ac.il \\ nardi@to.infn.it
}
\begin{abstract}
We present the predictions for hadron multiplicities in $pp$, $pA$ and $AA$ collisions at the LHC based on our approach to the Color Glass Condensate.
\end{abstract}
%\maketitle
%\section{Introduction: file preparation and submission}
We expect that at LHC energies, the dynamics of soft and semi-hard interactions will be dominated by parton saturation. In this short note we 
summarize our results for hadron multiplicities basing on the approach that we have developed and tested at RHIC energies in recent years \cite{KLN}; a detailed description of 
our predictions for the LHC energies can be found in Ref. \cite{KLN-LHC}. In addition, we will briefly discuss the properties of non-linear evolution at high energies, and their implications; details will be presented elsewhere \cite{KL}.  Our approach is based on the description of initial wave functions of colliding hadrons and nuclei as sheets of Color Glass Condensate. We use a corresponding ansatz for the unintegrated parton distributions, and compute the inclusive cross sections of parton production using $k_{\perp}$-factorization.  The hadronization is implemented through the local parton-hadron duality -- namely, we assume that the transformation of partons to hadrons is a soft process which does not change significantly the angular (and thus pseudo-rapidity) distribution  of the produced particles. Because of these assumptions, we do not expect our results be accurate for the transverse momentum distributions in AA collisions, but hope that our calculations (see Fig. 1a) will apply to the total multiplicities. 

\vskip0.3cm

While our approach has been extensively tested at RHIC,  an extrapolation of our calculations to the LHC energies requires a good theoretical control 
over the rapidity dependence of the saturation momentum $Q_s(y)$.  The non-linear parton evolution in QCD is a topic of vigorous theoretical investigations at present. Recently, we have investigated the role of longitudinal color fields in parton evolution at small $x$, and found that they lead to the following dependence of the saturation momentum on rapidity \cite{KL}:
  \beq \label{SOLSWWC}
  Q^2_s \lb Y \rb\,\,=\,\,\frac{ Q^2_s \lb Y =Y_0 \rb\,\,\exp\lb \frac{2\,\as}{\pi}\,\,\lb Y - Y_0 \rb \rb}{
  1\,\,+\,\,B\,Q^2_s \lb Y =Y_0 \rb\,\,\lb \exp\lb \frac{2\,\as}{\pi}\,\,\lb Y - Y_0 \rb \rb\,\,-\,\,1 \rb},
  \eeq
  where $B\,\,=\,\,1/(32\,\pi^2) \,\,(\pi\,R^2_A/{\as})$; 
  $R_A$ is the area of the nucleus, and $\as$ is the strong coupling constant. At moderate energies, Eq \eq{SOLSWWC} describes an exponential growth of the saturation momentum with rapidity; when extrapolated to the LHC energy this results in the corresponding growth of hadron multiplicity, see curve "1" in Fig.1b). At high energies, Eq \eq{SOLSWWC} predicts substantial slowing down of the evolution, which results in the decrease of hadron multiplicity as shown in Fig.1b) by the curve "2". In both cases, the growth of multiplicity is much slower than predicted in the conventional "soft plus hard" models, see Fig.1. We thus expect that the LHC data on hadron multiplicities will greatly advance the understanding of QCD in the strong color field regime. 
%%%%
\begin{figure}[ht]
%\begin{center}
  \begin{tabular}{cc} 
      \includegraphics[width=6.3cm]{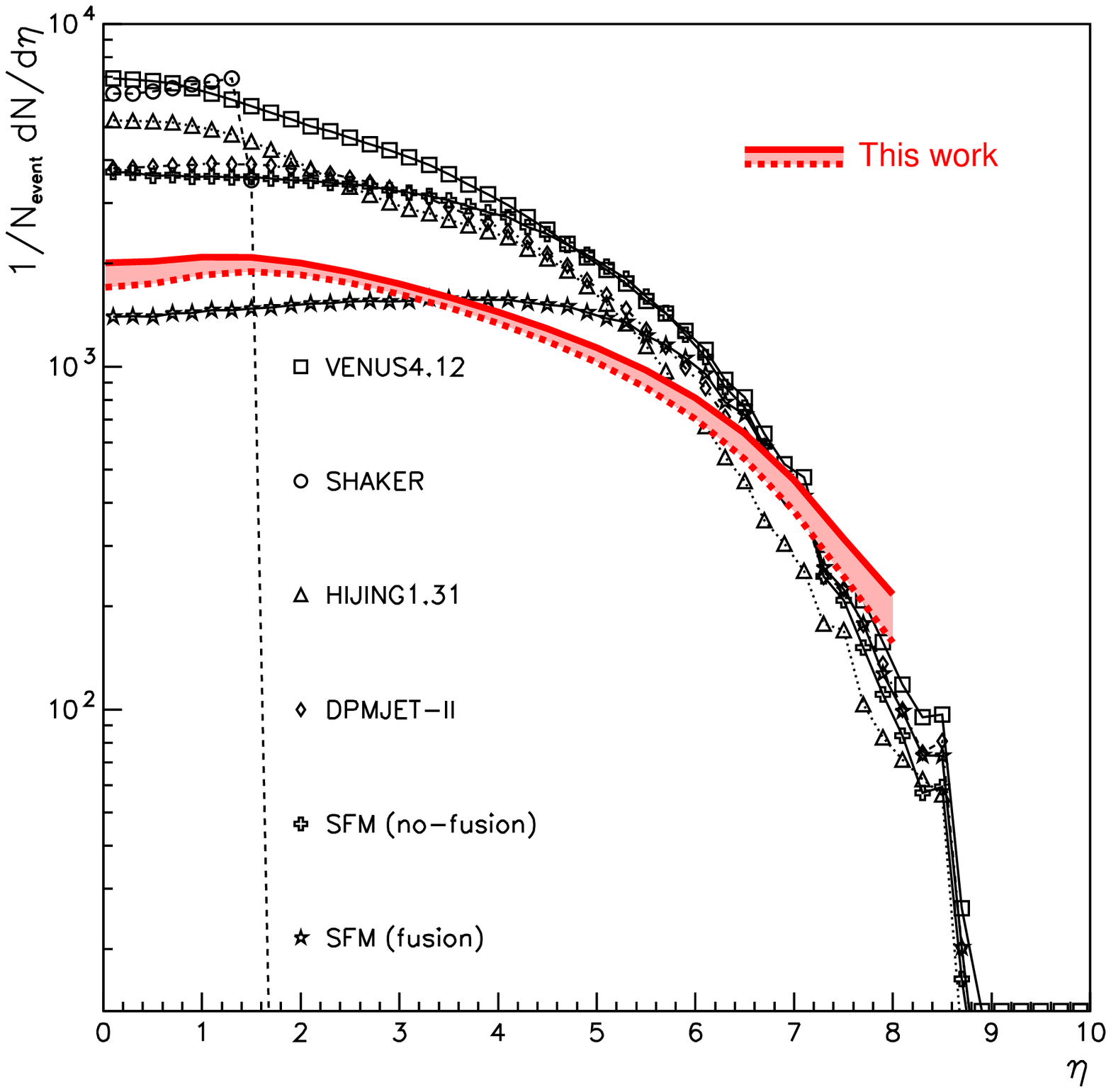} &
      \includegraphics[width=6.3cm]{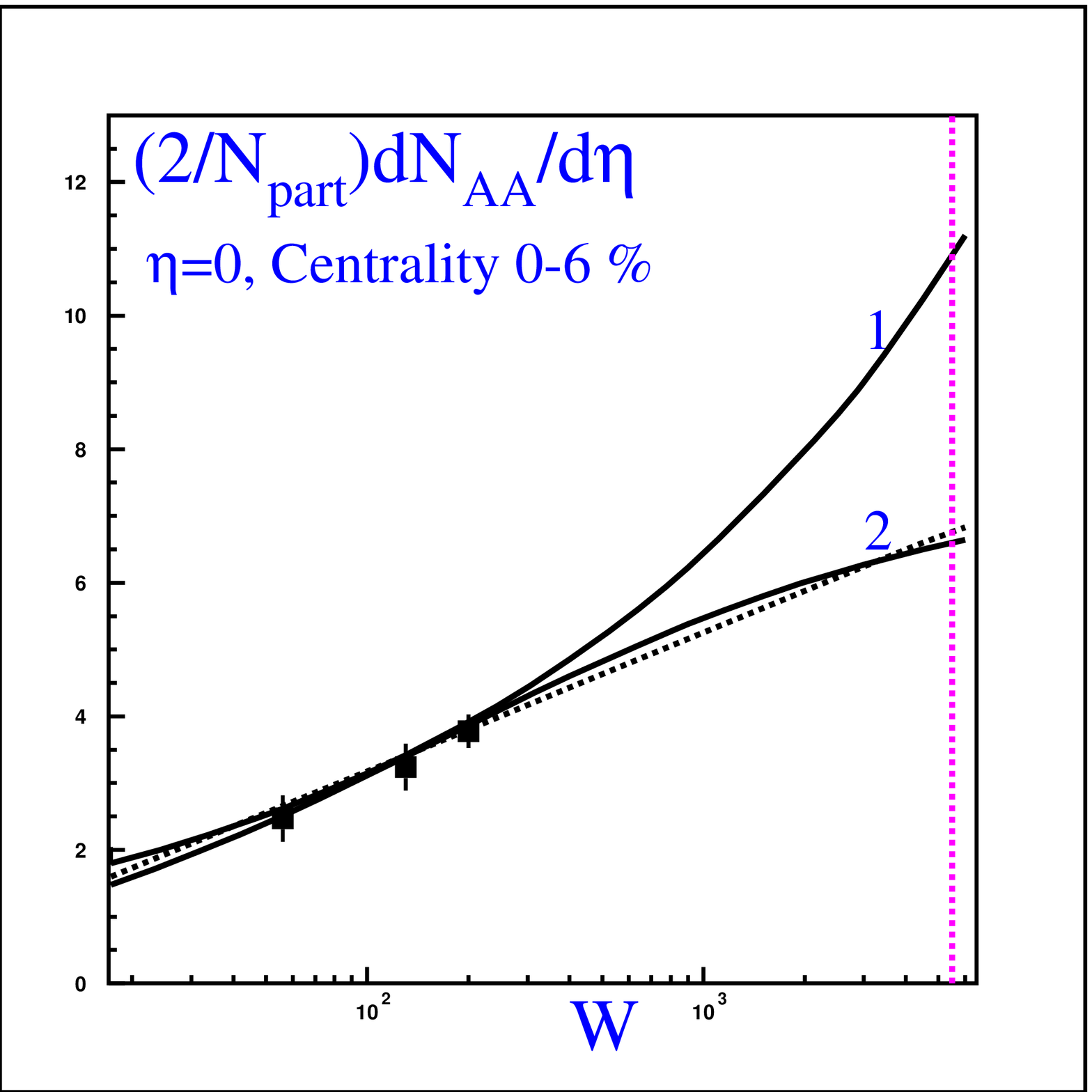}\\
         (a) & (b) 
\end{tabular}
%\end{center}
\caption{a) Charged hadron multiplicity in Pb-Pb collisions as a function of pseudo-rapidity at $\sqrt{s} = 5.5$ TeV; also shown are predictions from other approaches (from \cite{KLN-LHC}); b) Energy dependence of charged hadron multiplicity per participant pair in central AA collisions for different approaches to parton evolution (curves 1 and 2); also shown is the logarithmic fit, dashed curve (from \cite{KL}). }
\label{diag}
\end{figure}
%%%%%

The work of D.K. was supported by the U.S. Department of Energy under Contract No. DE-AC02-98CH10886. The work of E.L. was supported in part by the grant of Israeli Science Foundation founded by Israeli Academy of Science and Humanity.

\end{document}